\documentstyle{mn}
\def\lsim{\, \lower2truept\hbox{${<
\atop\hbox{\raise4truept\hbox{$\sim$}}}$}\,}

\def\gsim{\, \lower2truept\hbox{${>
\atop\hbox{\raise4truept\hbox{$\sim$}}}$}\,}

\input psfig.tex

\title{The dark matter distribution in disk galaxies}

\author[A. Borriello \& P. Salucci]
         {Annamaria Borriello$^1$ \& Paolo Salucci$^1$ \\
          (1) International School of Advanced Studies SISSA-ISAS -- Trieste, I}

\begin{document}

\maketitle

\begin{abstract}
We use high--quality  optical rotation curves  of 9 low--luminosity  disk
galaxies  to obtain the velocity  profiles  of the surrounding  dark matter
halos.  We find  that they increase linearly with radius at least out to  the
edge of the stellar disk,  implying that, over  the entire stellar  region,
 the density of the dark halo is about constant.

The properties of the  mass structure of these halos  are similar to
those found for  a number of dwarf and low surface brightness  galaxies, 
but provide a more  substantial evidence of   the discrepancy between  the
halo mass distribution  predicted in the Cold Dark Matter scenario and those
actually  detected  around  galaxies. We find that the density law
proposed by Burkert (1995)  reproduces the halo rotation curves,  with
halo central densities ($\rho_0 \sim$ 1--4 $\times 10^{-24}$ g cm$^{-3}$)
and  core radii ($r_0 \sim$ 5--15 kpc) scaling as  $\rho_0  \propto
r_0^{-2/3}$.

\end{abstract}


\section{Introduction}

Rotation curves (RCs) of disk galaxies are  the best  probe for  dark matter
(DM) on galactic scale. Notwithstanding the impressive amount of  knowledge
gathered in the past  20 years, some crucial aspects  of the  mass {\it
distribution} remain unclear. In fact, the  actual density  profile of dark 
halos  is still matter of debate; we do not even know  whether it is universal or
related to some galaxy  property, such as the total mass. This is partly
because such  issues are  intrinsically  crucial  and partly  because it is
often   believed that a  RC leads to   a quite  ambiguous information on the
dark halo   density distribution (e.g. van Albada et al. 1985).
However,  this  argument is true only  for rotation curves  of  low
spatial resolution,  i.e. with   no more  than 2  measures
per exponential  disk  length--scale $R_D$,  as is the case of  the
great majority of HI RCs. This is because  the galaxy  structure
parameters  are  very   sensitive to both the {\it amplitude } and the
{\it  shape} of the  rotation curve  in  the  region $0<r<R_D$, (e.g.
Blais-Ouellette et al., 1999)   which  corresponds  to  the region of the RC
steepest  rise.  No reliable mass model can be derived if such a  region
is  poorly  sampled and/or radio beam--biased. However, in  case  of
high--quality {\it  optical} RCs with  tens of independent  measurements in
the  critical region, the kinematics can probe  the  halo  mass distribution
and resolve their structure. 

\begin{table}
\begin{center}
\begin{tabular}{lccc}
\hline
\hline
\noalign{\smallskip}
Galaxy & R$_{D}$ (kpc) & V$_{opt}$ (km s$^{-1})$  &  M$_I$  \\
\noalign{\smallskip}
\hline
\noalign{\smallskip}
116-G12          &    1.7       & 133.5  &      $-20.0$  \\
531-G22          &    3.3       & 171.1  &      $-21.4$  \\
533-G4           &    2.7       & 151.1  &      $-20.7$  \\
545-G5           &    2.4       & 124.4  &      $-20.4$  \\
563-G14          &    2.0       & 148.9  &      $-20.5$  \\
79-G14           &    3.9       & 167.1  &      $-21.4$  \\
M-3-1042         &    1.5       & 148.0  &      $-20.1$  \\
N7339            &    1.5       & 172.7  &      $-20.6$  \\
N755             &    1.5       & 102.4  &      $-20.1$  \\
\noalign{\smallskip}
\hline
\end{tabular}
\end{center}

\caption{Observational properties of the sample galaxies:
(1)  Galaxy name. (2)  Length--scale of the exponential thin
disk (from Persic \& Salucci 1995).
The optical radius $R_{opt}$ is the radius encompassing
83\% of the light:  $R_{opt}=3.2 R_D$. (3)  Circular velocity at $R_{opt}$.
(4) Total $I$--band absolute magnitude (Mathewson, Ford \& Buchhorn, 1992).}

\end{table}

Since the dark component can be better traced  when  the disk contributes
to the dynamics in a modest way, it is convenient  to  investigate
DM--dominated objects, like dwarf and low surface brightness (LSB) galaxies.
It is well known that   for the latter there are strong   claims  of  dark
matter distributions  with cores of  constant density,   in disagreement
with the   steeply cusped  density  distributions of the Cold Dark Matter
Scenario. (Flores \& Primack, 1994; Moore, 1994; Burkert, 1995; Burkert \&
Silk, 1997; Kravtsov et al., 1998; McGaugh \& de Blok, 1998; Stil, 1999).
However,  these  findings  are  {\it 1)}  under the {\it caveat} that
the  low  spatial  resolution of the analysed RCs  does not bias
the mass modeling and  {\it 2)} uncertain, due to the limited amount of
available kinematical  data  (see van den Bosch et al., 1999).

\begin{figure*} 
\centerline{\psfig{file=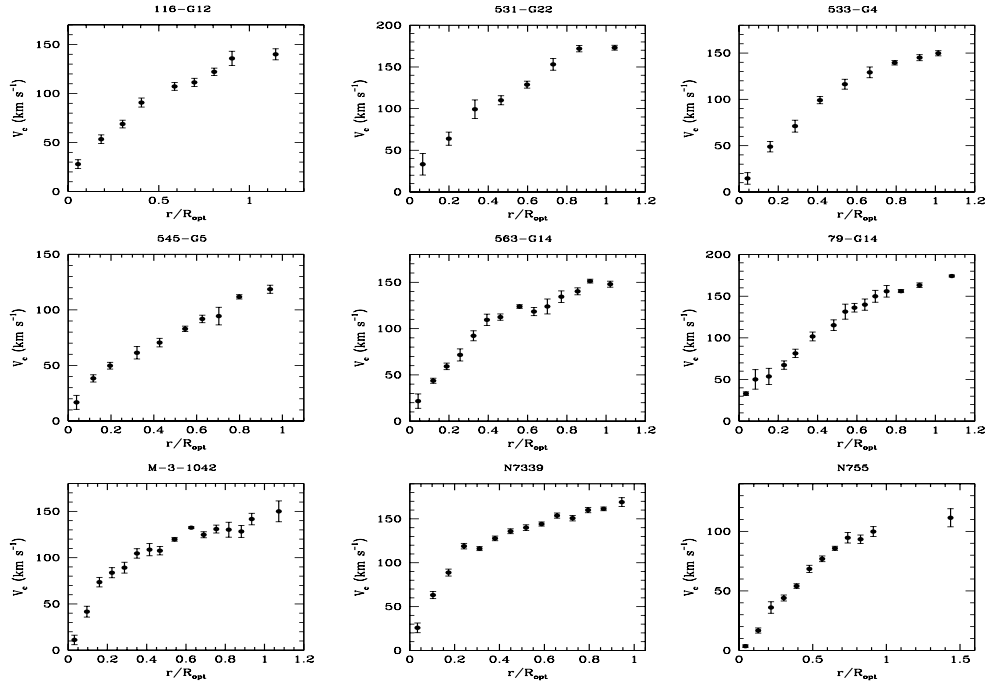,width=180mm,height=180mm}} 
\vspace{-7truecm}
\caption{Rotation curves of the  sample galaxies.} 
\end{figure*} 

In this paper we will investigate  the above--discussed  issue by  analysing
a number of high--quality {\it optical} rotation curves
of {\it low luminosity }  late--type spirals taken
from Persic and Salucci (1995, PS95), with  I--band absolute
magnitudes $-21.4<M_I<-20.0$ that, in terms of rotational velocities,
translates into $100 <V_{opt}< 170$ km s$^{-1}$. Objects in this
luminosity/velocity range are DM dominated (e.g. Persic, Salucci \& Stel,
1996), but their RCs, measured at the PS95 angular resolution of $2^{\prime
\prime}$, have a spatial resolution of  $w\sim 100 (D/10 \  {\rm Mpc})$
pc  and  $n_{data}\sim R_{opt}/w$ independent measurements.
For  nearby galaxies: $ w<< R_{D}$ and $n_{data}>25$.
Moreover, we select  rotation curves of bulge--less systems,  so  that  the
stellar disk  is the only baryonic component for  $r  \lsim  R_D$.

Since most of the properties of cosmological halos are claimed universal,
it is worth to  concentrate on a small and particular
sample of RCs, that  however   can  provide crucial information
on the  dark halo density  distribution. The  systematics and
the  cosmic variance  of the DM halos  are investigated
elsewhere (Salucci and Burkert, 1999; Salucci, 1999).   Finally, let us stress
that to estabilish the {\it actual} theoretical properties of CDM halos  or to
investigate {\it non--standard} CDM scenarios possibly  in agreement with
observations is beyond the scope of this work.

In \S2 we describe our sample of RCs.  We present  our mass modeling
technique and its results in \S3. In \S4 we discuss  the inferred halo
profiles and their properties. In \S5 we perform a disk--halo
rotation curve modeling by adopting a CDM density profile for
the dark halo. We summarize our results in \S6.
Throughout this paper we adopt a Hubble constant of
$H_0=75$ km s$^{-1}$ Mpc$^{-1}$.

\begin{figure*} 
\centerline{\psfig{file=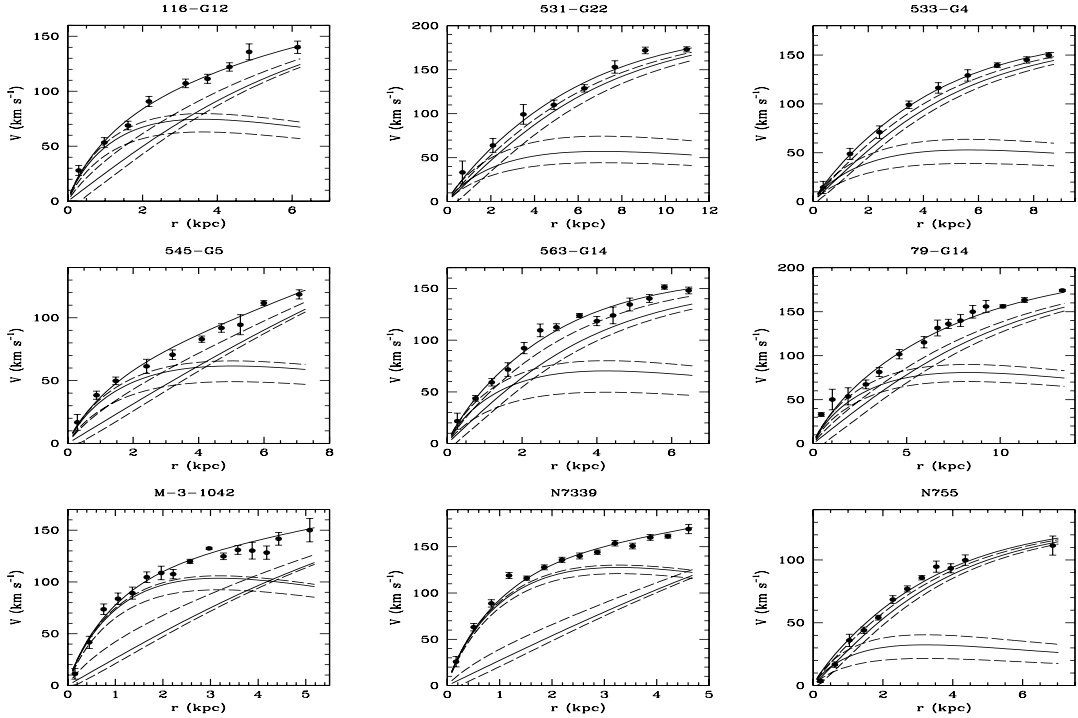,width=200mm,height=190mm}} 
\vspace {-7.6truecm} 
\caption{CDR model fits ({\it thick solid line})  to the RCs 
({\it points with errorbars}). Thin solid lines represent  the 
disk and  halo contributions. The  maximum disk and the minimum 
disk solutions are also plotted ({\it dashed lines}).} 
\end{figure*}

\section{Rotation Curves}

The  rotation curves of  the  Persic \& Salucci (PS95) `excellent' subsample
of $80$ galaxies are  all suitable  for an accurate mass modeling. In fact,
these RCs properly  trace the gravitational potential in that:  {\it 1)}
data extend at least to the optical radius, {\it 2)} they  are smooth and
symmetric,  {\it 3)} they have small {\it rms}, {\it 4)} they  have high
spatial  resolution  and a homogeneous radial data coverage,  i.e. about
$30-100$ data points homogeneously distributed with radius and between the
two arms. From this subsample we extract 9 rotation curves of low
luminosity galaxies  ($5 \times 10^9 L_{\odot} <L_I< 2 \times
10^{10} L_{\odot}$;  $100<V_{opt}< 170$ km s$^{-1}$), with their
$I$--band surface luminosity  being an (almost)  perfect  radial
exponential.    These two last criteria, not indispensable  to
perform the   {\it mass} decomposition, are however required to
minimize   the uncertainties of the resulting dark halo {\it density}
distribution. The selected RCs are shown in Figure 1 (for all  details we
refer to PS95). They are  still growing at $R_{opt}$, in that mostly tracing
the dark halo component.  Each RC has $7-15$  velocity points inside
$R_{opt}$,  each one being the average of $2-6$ independent data. The RC
spatial resolution  is better than $1/20\ R_{opt}$, the velocity  {\it rms}
is about $3\%$ and the RCs  logarithmic derivative  is generally known within
about 0.05.

\section{Mass modeling} 
 
We model the mass distribution as the sum of two  components: a stellar  disk
and a spherical dark halo.   By assuming  centrifugal equilibrium under the
action of the gravitational potential,  the  observed circular velocity can be
split into these two  components: 

\begin{equation} 
V^2(r)=V^2_D(r)+V^2_H(r) 
\end{equation}

\begin{table}

\begin{center} 
\begin{tabular}{lccc} 
\hline
\noalign{\smallskip}
Galaxy &  $\beta$ & \ \ \ \ $a$ & M$_{D}/$L$_I$  \\  
\noalign{\smallskip}
\hline
\noalign{\smallskip}

116-G12  &  $0.28^{+0.04}_{-0.08}$  &  $1.4^{+0.3}_{-0.2}$  & 
$1.0^{+0.1}_{-0.3}$  \\  

531-G22  &  $0.10^{+0.07}_{-0.04}$  &  $0.8^{+0.8}_{-0.2}$  & 
$0.3^{+0.2}_{-0.1}$  \\ 

533-G4   &  $0.11^{+0.05}_{-0.05}$  &  $0.8^{+0.2}_{-0.1}$  &   
$0.4^{+0.2}_{-0.2}$  \\ 

545-G5   &  $0.22^{+0.03}_{-0.08}$  &  $2.5^{+0.5}_{-0.5}$  &   
$0.7^{+0.1}_{-0.2}$  \\ 

563-G14  &  $0.20^{+0.06}_{-0.10}$  &  $0.8^{+0.2}_{-0.2}$  &  
$0.7^{+0.2}_{-0.3}$  \\ 

79-G14   &  $0.21^{+0.05}_{-0.05}$  &  $1.0^{+0.1}_{-0.1}$ &   
$0.7^{+0.2}_{-0.2}$   \\ 

M-3-1042 &  $0.44^{+0.02}_{-0.09}$  &  $1.9^{+0.1}_{-1.1}$  &  
$1.6^{+0.1}_{-0.3}$  \\ 

N7339    &  $0.49^{+0.02}_{-0.05}$  &  $2.4^{+1.0}_{-0.7}$  &   
$1.6^{+0.1}_{-0.2}$     \\   
N755     &  $0.09^{+0.05}_{-0.05}$  &  $1.0^{+0.1}_{-0.2}$  &   
$0.2^{+0.1}_{-0.1}$  \\  
 \noalign{\smallskip}
\noalign{\smallskip}
\hline 
\hline 
\end{tabular} 
\end{center} 
\caption {CDR mass models: (1)  Galaxy name; (2)-(3) Parameters of the
best--fit  CDR models with  their  $1\sigma$ uncertainties 
(\S3); (4) Disk mass--to--light ratio  in the $I$--band in solar units.} 
\end{table} 

This is true as long as one assumes the galactic disk be in dynamic 
equilibrium. By selection, the objects are bulge--less and the stellar 
component is  distributed like an exponential  thin disk. Light traces  the 
mass via     an assumed  radially  constant mass--to--light ratio.

In the r.h.s of  eq.(1) we neglect  the gas contribution $V_{gas}(r)$ since 
in  normal spirals  it  is usually   modest within the optical region (Rhee (1996), 
Fig. 4.13): $ \beta_{gas}\equiv  (V^2_{\rm gas}/V^2)_{R_{opt}} \sim 0.1$. 
Furthermore, high resolution HI  observations show that in  galaxies with  
RCs  similar  to those of the present  sample ( M33: Corbelli
\& Salucci, 1999; NGC300: Puche, Carignan \& Bosma, 1990;  N5585:
Cote \& Carignan, 1991; N3949 and N3917: Verheijen,  1997)  $V_{gas}(r)$ is
well represented by  $V_{gas}(r) \simeq 0$ for $r < R_D$ and: 

\begin{equation} 
V_{gas}(r) \simeq (20 \pm 5) (r-R_D) / 2 R_D \ \ \ \ \ \ \ \ \ \ \ \  
R_D \leq r \leq 3R_D  
\end{equation}

Thus, in the optical region:  {\it i)}  $V_{gas}^2(r)<<V^2(r)$  and {\it ii)}
$d(V^2(r)-V^2_{gas}(r))/dr  \gsim 0$.  This last condition implies that by 
including  $V_{gas}$ in the  r.h.s. of  eq.(1)   the  halo velocity profiles
would result {\it steeper} and  then  the core radius in the  halo density  
{\it larger}.  Incidentally,  this is not the case  for dwarfs and LSBs:  most
of  their  kinematics is  affected by the HI disk  gravitational pull in such
a way that  neglecting it could bias the determination of the DM density. 

The circular velocity profile of the disk is (Freeman, 1970): 
 
\begin{equation} 
V_D^2(r)=V^2_{opt} \ \beta \ \frac{r^2}{R^2_{opt}}\ 
\frac{(I_0K_0-I_1K_1)_{1.6r/R_{opt}}}{(I_0K_0-I_1K_1)_{1.6}} 
\end{equation} 
 
where $I_n$ an $K_n$ are the modified Bessel functions, $V_{opt}$ is the 
measured circular velocity at $R_{opt}$ and $\beta\equiv \left 
(\frac{V^2_D}{V^2}\right )_{R_{opt}}$.

The parameter $\beta$ 
represents the disk contribution to the total circular velocity at $R_{opt}$ 
and  can  vary   from 0 to 1. On grounds of simplicity, we choose $\beta$ as the disk 
free parameter  rather than the disk mass--to--light ratio. 

For the DM halo we  assume a spherical distribution,  whose contribution to 
the circular velocity $V_H(r)$ is given by (Persic, Salucci \& Stel 1996, 
Salucci  1997): 
 
\begin{equation} 
V_H^2(r)=V^2_{opt} \ \gamma \ (1+a^2) \frac{x^2}{(x^2+a^2)} 
\end{equation} 
 
where $x \equiv r/R_{opt}$ and $a$ is the core radius measured in units of 
$R_{opt}$. From eq.(1): $\gamma = (1-\beta)$.  Since we  normalize (at
$R_{opt}$)  the  velocity model $(V_H^2+V^2_D)^{1/2}$ to the observed rotation
speed   $V_{opt}$,  $\beta$ enters   explicitly in the halo velocity model 
and this  reduces  the free parameters  of the mass model to two.

It is important to remark that, out to $R_{opt}$, the  proposed Constant
Density Region (CDR) model  of eq.(4) is  neutral with respect to  the 
competing  mass models. Indeed,  by varying  $\beta$ and $a$, it 
approximately  reproduces  the  maximum--disk,  the  solid--body, the
no--halo, the  all--halo, the  CDM and the coreless--halo models.  For
instance, CDM halos  with  concentration parameter $c=5$  and $r_s=R_{opt}$
(see \S 5) are well fit by eq.(4) with $a \simeq  0.33$.

For each galaxy, we determine the values of the parameters $\beta $ and $a$
by means of a $\chi ^2$--minimization fit to the observed rotation curves:

\begin{equation}
V^2_{model}(r; \beta , a) = V^2_D (r; \beta) + V^2_H (r; \beta , a)
\end{equation}

A central role in  discriminating among the different mass decompositions  is 
played  by the derivative of the velocity field $dV/dr$.  It has been shown 
(e.g. Persic \& Salucci  1990b, Persic \& Salucci 1992) that by taking  into 
account the logarithmic gradient of the circular velocity field defined as:

\begin{equation} 
\nabla (r)\equiv \frac{d \log V(r)}{d \log r} 
\end{equation}

\begin{figure} 
\centerline{\psfig{file=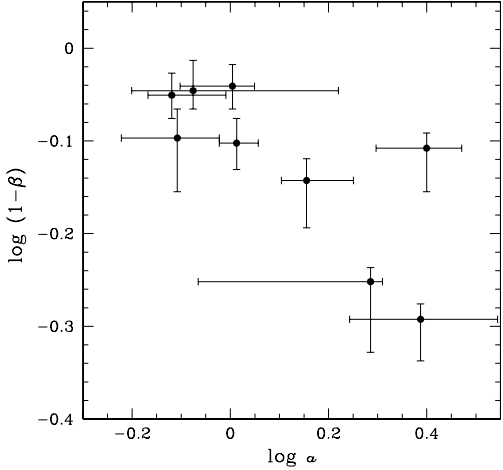,width=70mm}} 
\vspace{-0.5 truecm}
\caption{Relation between the halo parameters, ($a$ is in units 
of $R_{opt}$)} 
\end{figure}

one can significantly increase the amount of information available from 
kinematics and stored in the  shape of the rotation curve. 
So we consider   $\chi^2$-s   calculated on both velocities and logarithmic 
gradients:

\begin{equation} 
\chi^2_V =\sum^{n_V}_{i=1}\frac{V_i-V_{model}(r_i; \beta,a)} 
{\delta V_i} 
\end{equation} 
 
\begin{equation} 
\chi^2_{\nabla} = \sum^{n_{\nabla}}_{i=1}\frac{\nabla( 
r_i)-\nabla_{model}(r_i; \beta,a)}{\delta \nabla_i} 
\end{equation} 
 
where   $\nabla_{model} (r_i,  \beta,a)$ is computed from eq.$(3)-(4)$ 
and eq.(6).  As the linear combination of $\chi^2$-s \ still follows the 
$\chi^2$--statistics (Bevington \& Robinson 1992), we  derived the parameters 
of  the mass  models by  minimizing  a total $\chi^2_{tot}$, defined as: 
 
\begin{equation} 
\chi^2_{tot} \equiv \chi^2_V+ \chi^2_{\nabla} 
\end{equation} 

\begin{figure} 
\centerline{\psfig{file=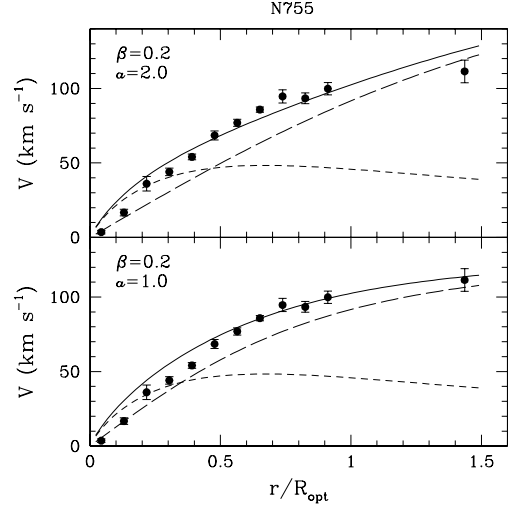,width=70mm}} 
\caption{N755 rotation curve compared with over-maximum-disk 
 models ($\beta=0.2$ and $a=2.0$ ({\it top}) or $a=1.0$ 
({\it bottom})). Solid line represents the total circular velocity, short-- and 
long--dashed lines represent the disk and halo contributions. The best--fit 
model for this galaxy has  $\beta=0.09^{+0.05}_{-0.05}$ and $a=1.0^{+0.1}_{-0.2}$, 
corresponding to $M_D/L_I = 0.2 \ M_{\odot}/L_{I\ \odot}$. We realize that 
 models  with  significantly higher disk mass--to--light ratio 
are inconsistent with the inner rotation curve, regardless of the halo
profile.}  
\end{figure}
 
The  above is the  $\alpha=1$ case of  the general relation $\chi^2_{tot} \equiv 
\chi^2_V+ \alpha  \cdot  \chi^2_{\nabla}$ we adopt in that  the 
circular  velocity at radius $r_i$  and the  corresponding  log-gradient 
$\nabla(r_i)$  are statistically independent. Notice  that   the 
$\chi^2_{tot}$  best--fit solutions  are not statistically  different from
those  obtained with the usual $\chi^2$ procedure,  however,  the 
ellipse uncertainty is  now remarkably reduced.

\begin{figure*}  
\vspace {-2.5truecm} 
\centerline{\psfig{file=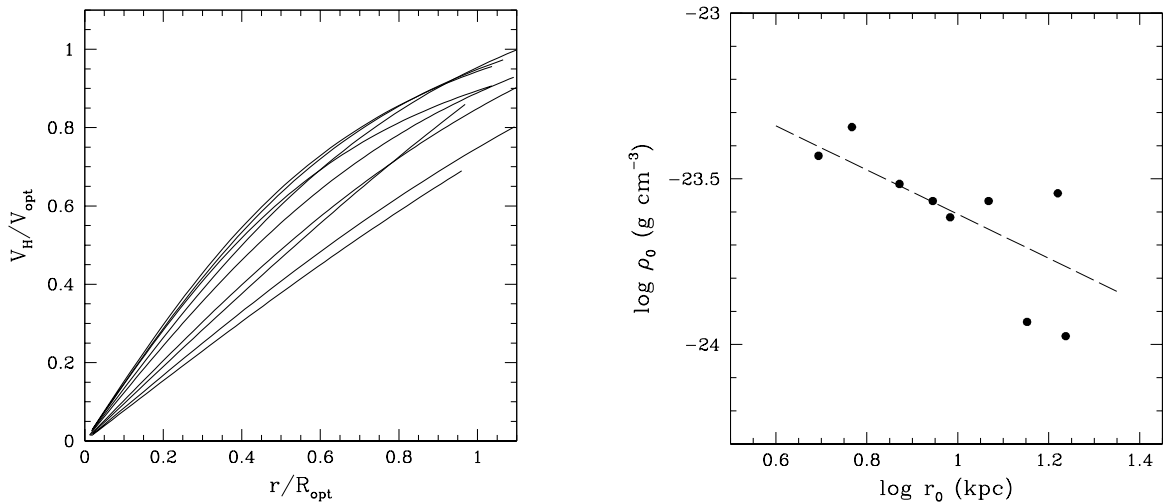,width=220mm}} 
\vspace {-20.6truecm} 
\caption{{\it (left)}:  The halo velocity profiles of the sample  galaxies. 
$V_H(r)$ rises almost linearly with radius: the DM halo  density remains 
approximately  constant {\it (right)}: halo central density $\rho_0$ {\it vs.} 
core radius $r_0$. The dashed line is the Burkert relationship  for  dwarf 
galaxies.} 
\end{figure*}

The parameters of the best--fit models 
are listed in Table 2, along with their $1\sigma$ uncertainties.  The 
mass models are well specified for each object: the  allowed  values 
for $\beta$ and $a$ span a small and continuous region of the ($a$, $\beta$) 
space.  We get a ``lowest" and a ``highest"  halo velocity 
curve by subtracting from $V(r)$  the maximum and the minimum disk 
contributions  $V_D(r)$  obtained by substituting in eq.(3) the parameter
$\beta$ with  $\beta_{best}+\delta\beta$ and $\beta_{best}-\delta\beta$,
respectively.

The derived mass models  are shown in  Figure 2, alongside with the separate 
disk and halo contributions. It is then obvious that the halo curve is 
steadily increasing,  almost linearly, out to the last data point. 
The disk--contribution $\beta $ and the halo core radius $a$  span a range 
from 0.1 to 0.5 and  from 0.8 to 2.5,  respectively. In each object  the 
uniqueness of the  resulting  halo velocity model   can be  realized by the 
fact that the maximum--disk and minimum--disk  models  almost coincide. 
In Figure 3  we show the correlation between 
the halo parameters:  halos which are more dynamically important  at 
$R_{opt}$ (i.e.  with higher $1-\beta$) have  smaller core radii.  Part of the 
 scatter of the relation may arise because the sample is limited in statistics 
and in  luminosity range (see Salucci and Burkert, 1999). 
Remarkably, we find that the size of the  halo density core is 
always greater than the disk characteristic scale--length 
$R_D$ and it  can extend  beyond  the disk edge (and the region 
investigated).

As regards to the HI disk  contribution, we  checked the consequence 
of neglecting it. Its contribution to the  circular velocitiy  (see \S 3,
Corbelli and Salucci,  1999) is computed and then  actually subtracted from 
$V(r)$. As result,  the  values of the parameter $\beta$ do not change whereas
we obtain   slightly larger  values of the core radius $a$. As a typical
example:   for N755,  considering  the gas distribution we get:  $\beta = 
0.05^{+0.07}_{-0.05}$ and $a=1.1^{+0.2}_{-0.2}$ to be compared with the 
value in Table 2.

Although the present sample is small for a thorough investigation of the 
stellar mass--to--light ratios, it is worth noticing that  the  disk
mass--to--light ratios  found    ($<M_D/L_I>\sim 0.8  M_{\odot}/L_{I \odot}$)
are  typical of late--type spirals  with young  dominant stellar
populations and ongoing star formation (e.g. de Jong, 1996),  which  adds to 
variations in stellar populations due to  differences in age,  metallicity, 
star formation history and to uncertainties in the estimate of 
distances and/or internal  extinctions 
In detail, we  ensured that for the  three galaxies with 
the  lowest disk mass--to--light 
ratios, a significantly larger value  of $\beta$ 
was  inconsistent with the  (inner)  rotation curve (see Figure  4).

Finally, it is worth stressing that the  present analysis computes the 
mass--to--light ratio from the model  parameters as a  secondary quantity; the 
uncertainties in Col.(4) of  Table 2 do not indicate 
the goodness of the {\it halo} mass models, but only specify how 
well we know this quantity.

\section{Dark Halos Properties}

In Figure 5 ({\it left})  we show  the  halo velocity profiles
for the nine galaxies. The halo circular velocities are  normalized to their
values  at $R_{opt}$ and  expressed  as a function of the normalized radius
$r/R_{opt}$. These normalizations  allow a  meaningful comparison between
halos of different masses: the radius scaling  removes the intrinsic dependence
of size on mass (more massive halos are  bigger), whereas the velocity scaling
takes into account that more luminous  galaxies have higher circular
velocities.  It is then evident that the halo  circular velocity, in every
galaxy, rises almost linearly with radius, at least out to the disk edge:

\begin{equation}
V_H(r) \propto r \ \ \ \hskip 1truecm   \ \ \ \ \ \ 0.05R_{opt}
\lsim r \lsim R_{opt} 
\end{equation}

\begin{figure*} 
\vspace {-1.8truecm} 
\centerline{\psfig{file=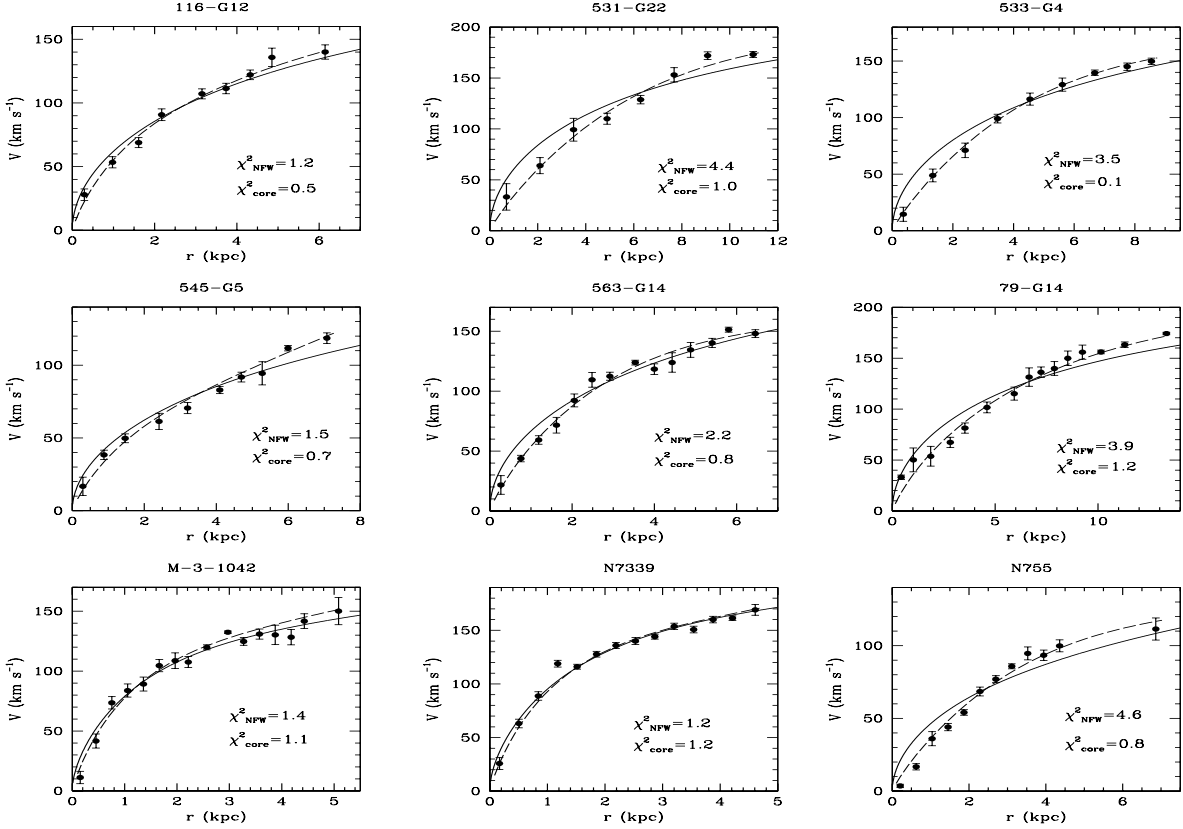,width=220mm,height=220mm}} 
\vspace {-8.3truecm} 
\caption{NFW  best--fits {\it solid  lines} of  the rotation curves 
{\it (filled circles)} compared with 
the CDR fits  {\it (dashed lines)}. The 
$\chi^2$ values are also indicated.} 
\end{figure*} 

The halo density profile has a well defined core radius within which
the density is approximately  constant. This is inconsistent with the
singular halo density  distribution  emerging  in the Cold Dark Matter (CDM)
scenario  of halo formation  (Navarro, Frenk \& White (NFW) 1995, 1996, 1997;
Cole \& Lacey, 1997; Tormen et al., 1997; Tissera \& Dominguez-Tenreiro, 1998;
Nusser \& Sheth, 1999). More precisely,  since the CDM
halos are, at small radii, likely  more cuspy than the   NFW profile:
$\rho_{CDM}\propto r^{-1.5}$ (Fukushige \& Makino, 1997; Moore et al., 1998;
Jing, 1999; Jing \& Suto, 1999; Ghigna et al., 1999), the  steepest CDM halo
velocity profile  $V_H(r) \propto r^{1/4}$ results   too shallow with
respect to  obervations and therefore inconsistent with eq.(10). 
Burkert (1995) has  proposed for the halo density distribution the following
phenomenological profile:
\begin{equation}
\rho_{B}(r)=\frac{\rho_0 r_0^3}{(r+r_0)(r^2+r_0^2)}
\end{equation}
where $\rho_0$ (the central density) and $r_0$ (the scale radius) are free
parameters. This density law has a core radius of size $r_0$,
and, at large radii,  converges to the NFW profile.
The two parameters   $\rho_0$ and $r_0$ are found  correlated: $\rho_0
\sim r_0^{-2/3}$, so that the  halo profiles reduce to a one--parameter family
of curves (Burkert, 1995).

We find that, by adjusting  $\rho_0$ and $r_0$,
we can  reproduce, over  the available radial range,  the halo velocity
profiles of  eq.(4) and Table 1. We show in Figure 5 ({\it right})  the pairs
($\rho_0$,  $r_0$) corresponding to the 9 objects. The  typical uncertainties
on $\rho_0$ and $r_0$ are estimated at a level of about  0.15 dex  and 0.07
dex, respectively: we then  confirm  the structure relation Burkert (1995)
found for 5 dwarf  galaxies (see Figure 5 {\it right}).

\begin{table*} 

\begin{tabular}{lccccccclc} 
\hline 
\hline 
\noalign{\smallskip}
\      & \multicolumn {4}{c}{$M_{\rm vir} \leq 2\cdot 10^{12} 
M_{\odot }$ } & \ &  \multicolumn {4} {c} {No limit on $M_{\rm vir}$} \\  
\noalign{\smallskip}
\hline 
\noalign{\smallskip}
Galaxy &  $\beta$ &  $c_{\rm vir}$ & $r_s$ (kpc) & M$_{D}/$L$_I$ & 
\   & $\beta$ &  $c_{\rm vir}$ & $r_s$ (kpc)  &  $\log (M_{\rm vir}/M_{\odot })$   \\ 
\noalign{\smallskip}
\hline 
\noalign{\smallskip}
116-G12  &   $0.01^{+0.04}_{-0.01}$ & $7.9^{+0.6}_{-0.4}$ & $26^{+4}_{-5}$ 
&    $0.04^{+0.14}_{-0.04}$ & \  \ & 
$0.01^{+0.04}_{-0.01}$ & $2.0^{+1.6}_{-0.2}$ & $456^{+50}_{-100}$ 
&  $14.3^{+0.3}_{-0.3}$   \\  
531-G22  & $0.01^{+0.03}_{-0.01}$ & $7.5^{+0.5}_{-0.5}$ & $28^{+4}_{-3}$ 
&    $0.03^{+0.10}_{-0.03}$ & \  \ & 
$0.01^{+0.01}_{-0.01}$ & $1.8^{+0.8}_{-0.1}$ & $472^{+20}_{-90}$ 
&  $14.0^{+0.2}_{-0.2}$    \\ 
533-G4   & $0.01^{+0.03}_{-0.01}$ & $7.0^{+0.5}_{-0.5}$ & $30^{+5}_{-4}$ & 
 $0.04^{+0.11}_{-0.04}$ &  \  \ & 
$0.01^{+0.03}_{-0.01}$ & $1.7^{+0.9}_{-0.2}$ & $500^{+20}_{-80}$ 
 &  $13.9^{+0.3}_{-0.3}$    \\  
545-G5    & $0.01^{+0.04}_{-0.01}$ & $4.7^{+0.3}_{-0.7}$ & $44^{+5}_{-4}$ & 
 $0.03^{+0.12}_{-0.03}$ &  \  \ & 
$0.01^{+0.02}_{-0.01}$ & $1.4^{+0.7}_{-0.1}$ & $479^{+30}_{-90}$ 
  &  $13.4^{+0.2}_{-0.3}$   \\  
563-G14   & $0.01^{+0.03}_{-0.01}$ & $8.7^{+0.3}_{-0.7}$ & $24^{+4}_{-4}$ & 
$0.03^{+0.10}_{-0.03}$ &  \  \ & 
$0.05^{+0.05}_{-0.05}$ & $2.0^{+1.6}_{-0.2}$ & $472^{+30}_{-70}$ 
&  $14.3^{+0.2}_{-0.3}$   \\ 
79-G14    & $0.01^{+0.04}_{-0.01}$ & $6.5^{+0.5}_{-0.5}$ & $32^{+3}_{-4}$ & 
 $0.03^{+0.14}_{-0.03}$ &  \  \ & 
$0.01^{+0.02}_{-0.01}$ & $1.6^{+0.4}_{-0.2}$ & $490^{+20}_{-110}$ 
&  $13.8^{+0.2}_{-0.3}$   \\ 
M-3-1042  & $0.20^{+0.05}_{-0.10}$ & $8.1^{+1.4}_{-0.6}$ & $26^{+3}_{-7}$ & 
 $0.72^{+0.18}_{-0.36}$ &  \  \ & 
$0.31^{+0.05}_{-0.06}$ & $1.7^{+1.3}_{-0.2}$ & $493^{+30}_{-120}$ 
&  $14.0^{+0.3}_{-0.5}$    \\  
N7339     & $0.18^{+0.02}_{-0.08}$ & $11.1^{+2.0}_{-0.6}$ & $19^{+2}_{-4}$ & 
$0.58^{+0.06}_{-0.26}$   & \  \ & 
 $0.38^{+0.04}_{-0.08}$ & $1.9^{+1.6}_{-0.3}$ & $497^{+20}_{-80}$ 
  &  $14.2^{+0.3}_{-0.4}$   \\ 
N755     & $0.03^{+0.02}_{-0.03}$ & $4.8^{+0.2}_{-0.8}$ & $43^{+3}_{-4}$ 
&  $0.05^{+0.04}_{-0.05}$   &  \  \ & 
$0.01^{+0.02}_{-0.01}$ & $1.5^{+0.9}_{-0.2}$ & $467^{+30}_{-110}$ 
 &  $13.6^{+0.1}_{-0.3}$   \\ 
\noalign{\smallskip}
\noalign{\smallskip}
\hline 
\hline 
\end{tabular} 

\caption{NFW mass models: (1) Galaxy name; (2)-(3)-(4)-(5)  Parameters of the
NFW best--fit models with their  $1\sigma$ uncertainties. An upper limit of $2
\cdot 10^{12}  M_{\odot}$ is imposed on $M_{\rm vir}$.  Disk mass--to--light
ratios  in the $I$--band in Col.(5) are in solar units;  (6)-(7)-(8)-(9) 
Parameters of the best--fit models  with no  constraint on the virial mass.} 

\end{table*}

Let us stress that the CDR  halo properties of  eq.(10)
raise important issues by themselves and  make quite irrelevant,
in  comparing  theory and observations,  arguments {\it per se}
important  such as the CDM halos cosmic variance, the actual value of the
concentration parameter or the effects of  baryonic infalls or outflows (e.g.
Martin, 1999; Gelato \& Sommer-Larsen, 1999).

We also want to remark that our finding  implies that disk galaxies are
embedded   (inside $R_{opt}$) in a {\it single}
dark halo. Indeed, if more dark halos were relevantly present, then, in order
to not violate the constraint given by eq.(10),
all of them should have the same solid body velocity
profile.

\section{The inadequacy of CDM mass models}

Although the mass models of eq.(4) converge to a distribution
with an inner core  rather than with a central spike, i.e. to
$a\simeq 1$ rather than to $ a \simeq 1/3$,
it is worth, given the importance of such  result,
to also check  in a direct way the (in)compatibility of the CDM
models  with the observed kinematics.
We assume the  two--parameters  functional form for the halo  density
by Navarro, Frenk \& White (NFW, 1995, 1996, 1997):

\begin{equation}
\rho_{\rm NFW}(r) = \frac{\rho_s}{(r/r_s)(1+r/r_s)^2}
\end{equation}

where $r_s$ is a characteristic inner radius and $\rho_s$ the
corresponding density. In order to translate the density profile into a
circular velocity curve for the halo, we make use of virial
parameters: the halo virial radius $R_{\rm vir}$,  defined as the radius
within which the mean density is $\Delta_{\rm vir}$ times the mean universal
density $\rho_m$ at that redshift, and the associated virial mass  $M_{\rm
vir}$ and velocity $V_{\rm vir} \equiv G M_{\rm vir} / R_{\rm vir}$. By
defining the concentration parameter as  $c_{\rm vir} \equiv R_{\rm vir}/r_s$
the halo circular velocity $V_{\rm CDM}(r)$ takes the  form (Bullock et al.
1999):

\begin{equation}
V_{\rm CDM}^2(r)= V_{\rm vir}^2 \frac{c_{\rm vir}}{A(c_{\rm vir})}
\frac {A(x)}{x}
\end{equation}

where $x\equiv r/r_s$ and $A(x)\equiv \ln (1+x) - \frac{x}{1+x}$.
As the relation between $V_{\rm vir}$ and $R_{\rm vir}$ is fully
specified by the background cosmology, we assume the currently
popular $\Lambda$CDM cosmological model,  with $\Omega_m = 0.3$,
$\Omega_{\Lambda} =0.7$ and $h=0.75$, in order to reduce from three to two
($c_{\rm vir}$ and $r_s$) the independent parameters characterizing the model.
According to this model, $\Delta_{\rm vir} \simeq 340$ at $z \simeq 0$.
The choice is conservative:  a high density $\Omega_m=1$ model,  with a
concentration parameter $c_{\rm vir}>12$,  is definitely  unable to
account for the observed galaxy kinematics (Moore 1994).

Though N--body simulations  and semi-analytic investigations indicate
that  the two parameters $c_{\rm vir}$ and $r_s$ seem to correlate,
we  leave them independent  to  increase  the chance of a good
fit. Since  the objects  in our sample  are of  low luminosity,
i.e. $L_*/12 \lsim  L_I \lsim  L_*/3$, we conservately  set for the
halo mass $M_{vir}$ an  upper limit of
$M_{up}= 2 \times 10^{12} M_\odot$, comparable to the total mass of
the  Galaxy (e.g. Wilkinson \& Evans, 1999) and to the mass of
bright spirals in pairs (e.g. Chengalur, Salpeter \& Terzian,
1993). This value is further justified by considering
that,  if  $M_{vir}>M_{up}$ in the above--specified luminosity range,  then,
the amount of dark matter locked in spiral galaxies $\Omega_S =
\int^{L_*/3}_{L_*/12} M_{vir}\ \phi(L)\ dL$, with $\phi(L)$ the galaxy
luminosity function, would much exceed  0.1, and would be  unacceptable for
the assumed cosmological model having  $\Omega_m=0.3$.

We performed the fit to the data with the
$\chi^2_{tot}$ minimization technique described in \S3;  the
results are reported in Table 3 , Col.(1)...(5). In Figure 6 we compare
the CDR and the NFW  models: for 7 galaxies the NFW model is
unacceptably worse than the CDR solution, whereas  for 2
objects (M--3--1042 and N7339) the goodness level of the two
different fits is comparable, but the virial mass in the case of
CDM is quite high:   $M_{vir}\sim 2 \times 10^{12} M_{\odot}$.
Moreover, the CDM models have extremely low value for the disk
mass--to--light ratio (see Table 3 Col.5). The majority  of the objects  have
$M_D/L_I \leq 0.05$, in obvious   disagreement  with the  spectro--photometric
properties of spirals,  that indicate  mass--to--light ratios at least 10
times higher.

The inadequacy of the CDM model for our sample galaxies  is even more
evident if one  performs the fit after removing any
constraint on virial mass. Indeed,  (results in Table 3, from Col.6 to 9),
good fits are obtained only  for very low values of the concentration
parameter ($c_{\rm vir} \simeq 2$) and for uncomfortably large virial velocities
and masses ($V_{\rm vir} \simeq 600-800 \ {\rm km \ s^{-1};}  M_{\rm vir}
\simeq 10^{13}-10^{14} M_{\odot}$). These results  can be explained
as effect of the  attempt of the minimization routine to fit  the  $V(r)
\propto r^{0.5}$ NFW velocity profile to data intrinsically linear in $r$.

\section{Summary and conclusions}

We have  performed the disk--halo decomposition for a  well suited sample of
9 bulge--less  disk galaxies  with $100\leq V_{opt } \leq 170$ km s$^{-1}$.
These galaxies  have a relevant amount  of dark matter:  the
contribution of the luminous matter to the dynamics  is  small and
it can be properly  taken into account. Moreover,  the high spatial resolution
of the available  rotation  curves  allows us to obtain  the separate  dark
and luminous density profiles.
We find that dark matter halos have  a constant central
density region whose size exceeds the stellar disk length--scale $R_D$.
These halo profiles disagree with the cuspy  density
distributions typical of CDM halos (e.g. Navarro, Frenk \& White, 1997;
Kravtsov et al., 1998), which, therefore,  fail to account for the  actual DM
velocity data. On the other hand, these  halo velocities are
well described in terms  of  the Burkert density  profile,
an empirical functional form whose two structure parameters (central density
and core radius) are  related through: $\rho_0 \sim~r_0^{-2/3}$.

This work is complementary to that of Salucci (2000) who  derived
for  140 objects of different luminosity $\nabla_h$, the logarithmic gradient of
the   halo velocity slope {\it at} $R_{opt}$.  Reminding that $\nabla_h=0$
and  $\nabla_h=1$ mean an isothermal and a solid-body regime, respectively,
let us stress that the result found:  $\nabla_h \leq 1$ completely
supports, at the optical edge, the results of the present paper obtained for
a small sample of spirals over the entire  stellar disk.

Finally, the  dark halo velocity linear rise   from $0.25 R_D$
to $\sim 3 R_D$ sets  a serious upper limit  to  the  dynamical  relevance
of  CDM--like dark halos in  spirals. Indeed, once we rule out a CDM halo,
also the claim  by Burkert and Silk (1997)
of {\it two} dark halos,  a MACHO dark halo with a CDR
profile and a standard  CDM halo,  meets a  difficulty.
Indeed, in this case, in order to satisfy eq.(10),
the MACHO halo should  account for $>95\%$ of the dark mass
inside  $R_{opt}$. Then,  it would dominate the dynamics out to
well beyond 25 $R_D$. The CDM halo would then  have  dynamical
importance in regions so external that its cosmological
role itself  would be in question.

\end{document}